\documentclass[
superscriptaddress,
 amsmath,amssymb,
 aps,
 prl,
 twocolumn,
floatfix
longbibliography
]{revtex4-2}

\usepackage[colorlinks,linkcolor=blue,anchorcolor=blue,citecolor=blue,urlcolor=blue]
{}
\usepackage{graphicx}
\usepackage{color}

\begin{document}

\date{\today}

\title{Triplet of kindred prompt-2$p$ emitters in mass-8 proton-rich nuclei}

\def\WUPHYS{Department of Physics, Washington University, St. Louis, Missouri 63130, USA.}
\def\FRIB{Facility for Rare Isotope Beams, Michigan State University, East Lansing, Michigan 48824, USA.}
\def\PAMSU{Department of Physics \& Astronomy, Michigan State University, East Lansing, Michigan 48824, USA.}

\def\Fudan{Key Laboratory of Nuclear Physics and Ion-beam Application (MOE), Institute of Modern Physics, Fudan University, Shanghai 200433, China.}
\def\Shanghai{Shanghai Research Center for Theoretical Nuclear Physics, NSFC and Fudan University, Shanghai 200438, China.}
\def\MSUPHYS{Department of Physics and Astronomy, Michigan State University, East Lansing, Michigan 48824, USA.}
\def\MSUCHEM{Department of Chemistry, Michigan State University, East Lansing, Michigan 48824, USA.}
\def\WUCHEM{Department of Chemistry, Washington University, St. Louis, Missouri 63130, USA.}
\def\ANL{Physics Division, Argonne National Laboratory, Argonne, IL 60439, USA.}
\def\WesternM{Department of Physics, Western Michigan University, Kalamazoo, Michigan 49008, USA.}
\def\Stores{Department of Physics, University of Connecticut, Storrs, Connecticut 06269, USA.}
\def\Lanzhou{Institute of Modern Physics, Chinese Academy of Sciences, Lanzhou 730000, China.}
\def\LSU{Department of Physics and Astronomy, Louisiana State University, Baton Rouge, Louisiana 70803, USA}

\author{R.J. Charity}
\affiliation{\WUCHEM}
\author{G.H. Sargsyan}
\affiliation{\FRIB}
\author{K.D. Launey}
\affiliation{\LSU}
\author {T.B. Webb}
\affiliation{\WUPHYS}
\author {K.W. Brown}
\affiliation{\FRIB}
\affiliation{\MSUCHEM}
\author {L.G. Sobotka}
\affiliation{\WUCHEM}
\affiliation{\WUPHYS}

\begin{abstract}
A triplet of kindred prompt-2$p$ emitters in $A$=8 nuclei has been demonstrated. Two of these are the ground state of $^8$C and its isobaric analog state in $^8$B, both of which are analogs of the halo or thick-skinned nucleus $^8$He.
The third member is the recently found fourth 1$^+$ state in $^8$B. This new $^8$B state at $E^*$=8.4 MeV was observed to decay to the ground state of $^6$Li by 2$p$ emission.
 Momentum correlations between the decay products indicate that it is not a sequential 2$p$ decay through a $^7$Be intermediate state, but indicative of prompt 2$p$ emission with correlations similar to those of the other members of the triplet. \textit{Ab initio} calculations with the symmetry-adapted no-core shell model indicate that these three states have very similar spatial wavefunctions, but with nucleons coupling to different spins, isospins, and isospin projections. The triplet of 2$p$ emitters all have oblate shapes and by stripping two protons equatorially, decay to states which have prolate shapes.

\end{abstract}

\maketitle

\begin{figure}[!htb]
\includegraphics[width=1\linewidth]{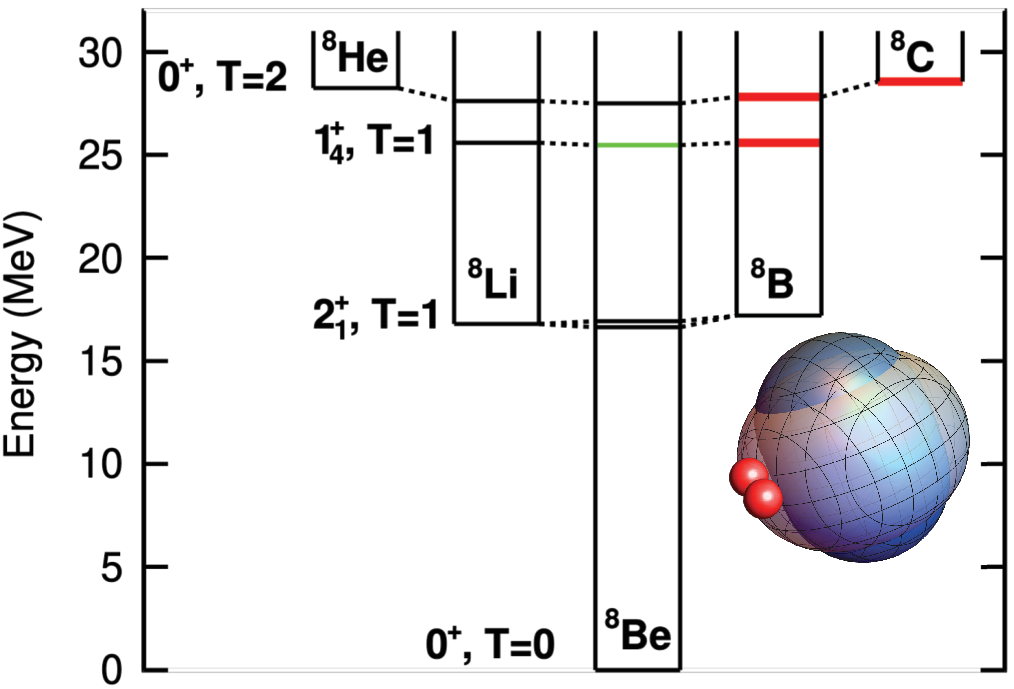}
\caption{An $A$=8 isobar diagram showing the states of interest in this work. The triplet of 2$p$ emitters are indicated by the thick red lines,
and schematically shown by their dominant shape (oblate) and the shape of the decay product (prolate). The analog of the 1$^+_4$ state in $^8$Be is unknown, but indicated by the green line at the expected location.} 
\label{fig:isoDia}
\end{figure}

As one goes away from the valley of stability on the nuclear chart one finds exotic nuclei that have neutron halos, on the neutron-rich side, and decay by the simultaneous emission of two protons, on the proton-rich side. These halo systems and the prompt 2$p$ emitters can sometimes be associated with similar nuclear structure. For instance, the mirror of the two-neutron-halo nucleus $^6$He is $^6$Be which is not a two-proton halo system as the two valence protons are unbound, but these two protons are emitted simultaneously \cite{Egorova:2012}. Another example is $^8$He, often described as having a  4$n$ halo \cite{AlKhalili:2004}, and its mirror $^8$C decays by two sequential steps of 2$p$ emission \cite{Charity:2010}. These two nuclei are the end members of an isospin multiplet which connects a series of 0$^+$, $T$=2 states with $A$=8 (see Fig.~\ref{fig:isoDia}) with similar exotic structure. 

Prompt 2$p$ decay is an exotic decay mode that is often found  for ground states of even-$Z$ isotopes just beyond the proton drip line \cite{Pfutzner:2023}.  
For the heavier ground-state 2$p$ emitters, this decay represents the only energy-conserving particle-decay mode. For lighter 2$p$ emitters, there is often an energetically-allowed 1$p$ decay process through an intermediate state which is  very broad (short lived) blurring the distinction between prompt and sequential 2$p$ decay. 

The isobaric analog states of the prompt-2$p$-decaying ground states of $^8$C and $^{12}$O, respectively in $^8$B (Fig.~\ref{fig:isoDia})
and $^{12}$N, also undergo prompt 2$p$ decay \cite{Charity:2010,Brown:2014,Webb:2019a}. 
However, these decays are also forced by conservation laws as they represent the only particle decay mode that conserves both energy and isospin. 

Examples of 2$p$ decay from excited states of isotopes inside the drip line where such restrictions from conservation laws are typically absent, are rare and must result from some peculiar nuclear structure favoring 2$p$ emission over 1$p$ decay. One such example is the second 0$^+$ state in $^{10}$C \cite{Charity:2022} which is the mirror of a highly-deformed $\alpha$-cluster configuration in $^{10}$Be \cite{Freer:2006}. In this case there is at present no theoretical understanding for the strength of the 2$p$ decay.  

In the present work, we find another case of 2$p$ decay with competing 1$p$ decay channels, specifically the 1$^+_4$ state in $^8$B, however in this case {\it ab initio} nuclear-structure calculations provide an explanation for the exotic decay.  This $^8$B state should be part of a  $J^\pi$=1$^+$, $T$=1  isospin multiplet  located in energy just below the $T$=2 multiplet (Fig.~\ref{fig:isoDia}).  These $T$=1 states have similar exotic structure to the $T$=2 states to which they are related largely by a spin-flip of one of the nucleons. The proton-rich members (red levels) of the $J^\pi$=0$^+$ $T$ = 2 and $J^\pi$=1$^+$ $T$=1 multiplets comprise a triplet of 2$p$ emitters with similar decay correlations.

\begin{figure}[!htb]
\includegraphics[width=0.9\linewidth]{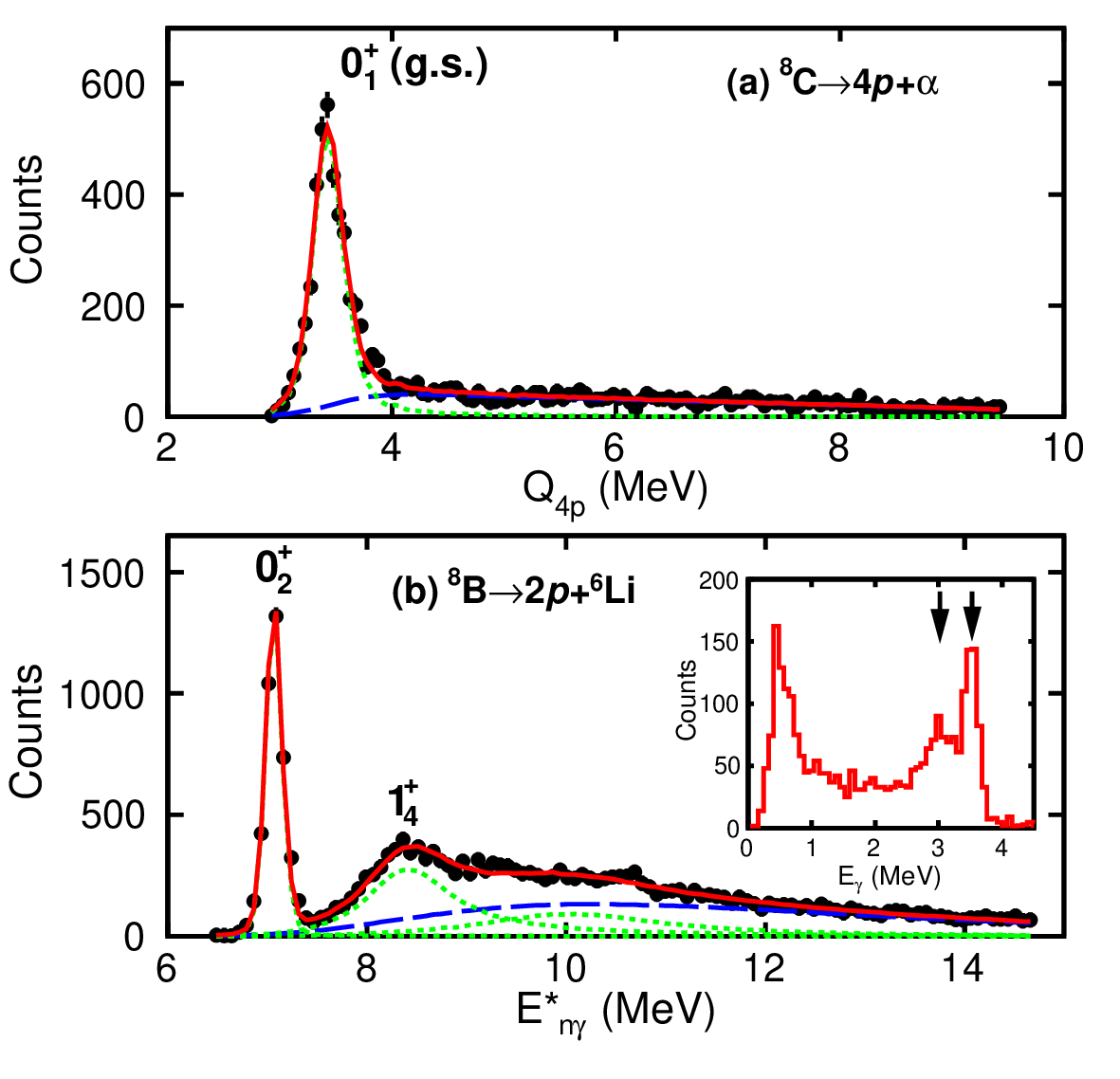}
\caption{Data points show invariant-mass measurements associated with 2$p$ emission in $A$=8 nuclei.
(a) The spectrum of 4$p$ decay energy for $^8$C states produced following the neutron knockout from a $^9$C beam \cite{Charity:2010}. (b) $^8$B excitation-energy distribution of 2$p$+$^6$Li events following proton knockout from this $^9$C beam \cite{CharityLongArxiv}. The solid red curves are fits to these data with the dashed blue curves showing the mix-event background \cite{Charity:2023,CharityLongArxiv} and the fitted peaks are shown as the dotted green curves. The inset in (b) shows the Doppler-corrected $\gamma$-ray energy distribution for events in the narrow 0$^+_2$ peak where the full-absorption and first-escape peaks (arrows) associated with the $\gamma$ decay of the 3.536-MeV 0$^+$ isobaric analog state in $^6$Li is observed.   
} 
\label{fig:Inv_2pLi6}
\end{figure}

{\it Experimental Data.---}
The 0$^+_1$ ground state of $^8$C was first observed in missing-mass spectroscopy \cite{Robertson:1974,Tribble:1976,Robertson:1976} and, from its measured mass, is found to be unstable for decay into the 4$p$+$\alpha$ exit channel. This exit channel was first investigated with invariant-mass spectroscopy where $^8$C resonances were created via neutron knockout from an $E/A$=69 MeV $^9$C beam by a $^9$Be target \cite{Charity:2010,Charity:2011}. The decay-energy spectrum from that work is shown in Fig.~\ref{fig:Inv_2pLi6}(a) where the narrow 0$^+_1$ ground-state peak sits upon little background. Correlations between the momentum vectors of the decay fragments indicate that this state decays by two steps of prompt 2$p$ emission \cite{Charity:2010,Charity:2011}. It is the first step, the prompt 2$p$ decay of $^8$C to the 0$^+_1$ ground state of $^6$Be, that is of importance for the present work.  

\begin{figure}[!htb]
\includegraphics[width=0.9\linewidth]{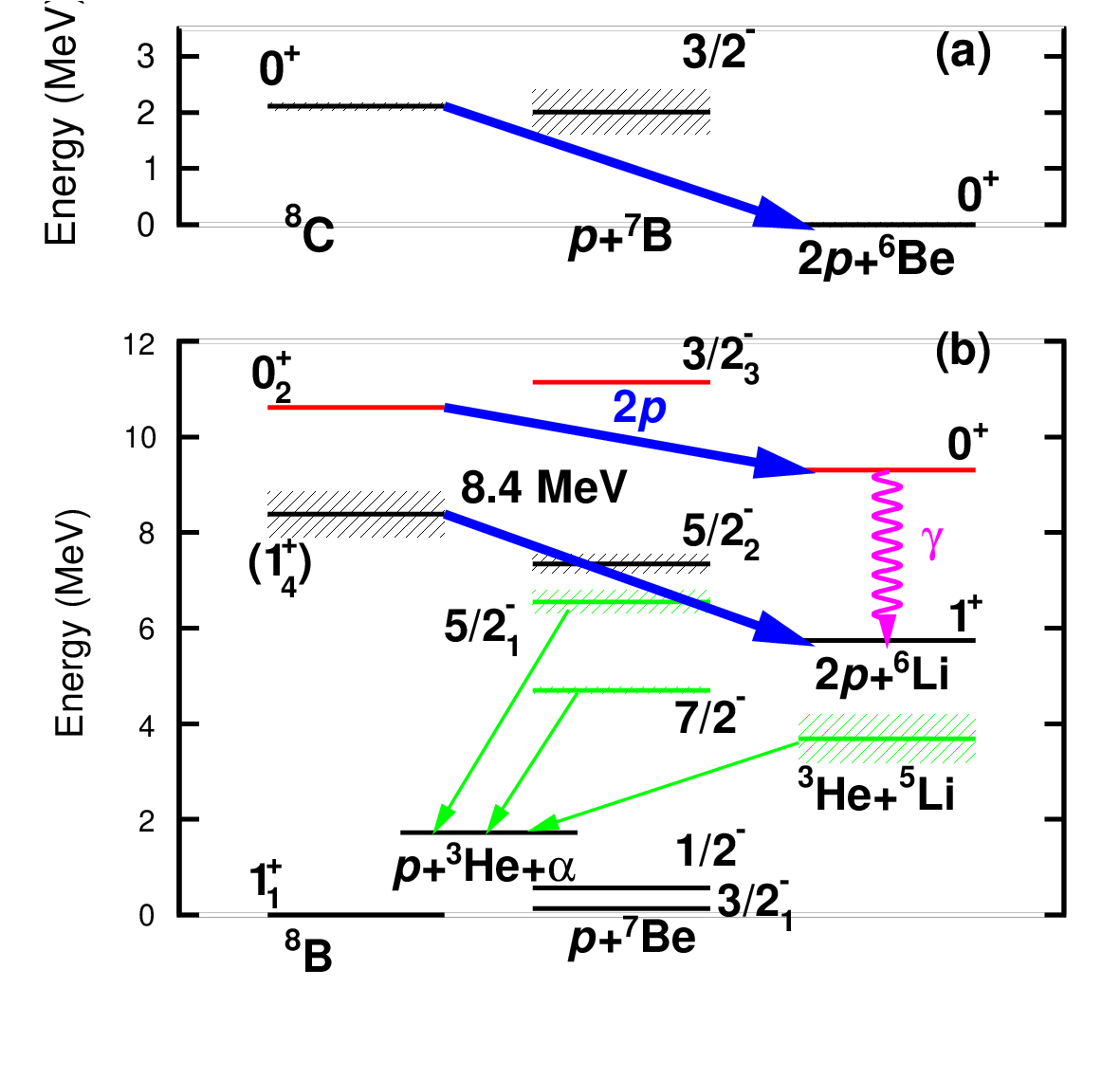}
\caption{Decay diagram for the 2$p$-emitting states in (a) $^8$C and (b) $^8$B.  The levels indicted by the red lines have high isospin and the green levels $\alpha$ decay. 
} 
\label{fig:level_2pLi6}
\end{figure}

With the same reaction, but knocking out a proton instead of a neutron, 2$p$ emission was also found to occur in $^8$B using invariant-mass spectroscopy of the 2$p$+$^6$Li exit channel \cite{Charity:2010,Brown:2014}. Figure~\ref{fig:Inv_2pLi6}(b) shows the $^8$B excitation energy calculated assuming that the decay populates $^6$Li in its 1$^+$ ground state. This assumption is incorrect for the narrow low-energy peak as the coincidence Doppler-corrected $\gamma$-ray energy spectrum, shown by the inset, shows the clear signature of the 3.562-MeV $\gamma$ ray (full absorption and first-escape peaks indicated by the two arrows) depopulating the 0$^+_1$ isobaric analog state in $^6$Li \cite{Brown:2014}. Thus the true excitation energy of this 2$p$-emitting state should be increased by the $\gamma$-ray energy. This makes the excitation energy of this state 10.619 MeV, the known excitation energy of 0$^+_2$, the isobaric analog of the ground state of $^8$C \cite{ENSDF}. The decays of the $T$=2 analog states [$^8$C(0$^+_1$) and $^8$B($0^+_2$)] are illustrated in Fig.~\ref{fig:level_2pLi6}. Both of these states 2$p$ decay to $T$=1 analog states [$^6$Be(0$^+_1$) and $^6$Li(0$^+_1$)].

In this work we consider the nature of the second peak in Fig.~\ref{fig:Inv_2pLi6}
at $E^*_{n\gamma}$=8.4 MeV. This peak is associated with a state of lower excitation energy than that for the narrow 0$^+_2$ peak as it decays to the ground state of $^6$Li \cite{CharityLongArxiv}, i.e. there is no coincidence $\gamma$ ray and thus $E^*$=$E^*_{n\gamma}$.
The existence of this state was confirmed from a second experiment, where the fragmentation products from an $E/A$=65-MeV $^{13}$O beam with a $^9$Be target \cite{CharityLongArxiv} are analyzed with the invariant-mass method. The weighted mean resonance parameters, from the two data sets, are $E^*$=8.40(4)~MeV and $\Gamma$=0.88(7)~MeV \cite{CharityLongArxiv}.

The level scheme for the decay of the 8.4-MeV state, shown in Fig.~\ref{fig:level_2pLi6}(b), indicates the final products can be produced through either one step of  prompt 2$p$ decay step or two sequential steps of 1$p$ decay.

For the latter, there is only one known proton-decaying state in $^7$Be (5/2$^-_2$) through which the sequential decay could proceed. 

Information on the nature of the 2$p$ decay can be gleaned from the momentum correlations between decay fragments. 

\begin{figure}[!htb]
\includegraphics[width=0.9\linewidth]{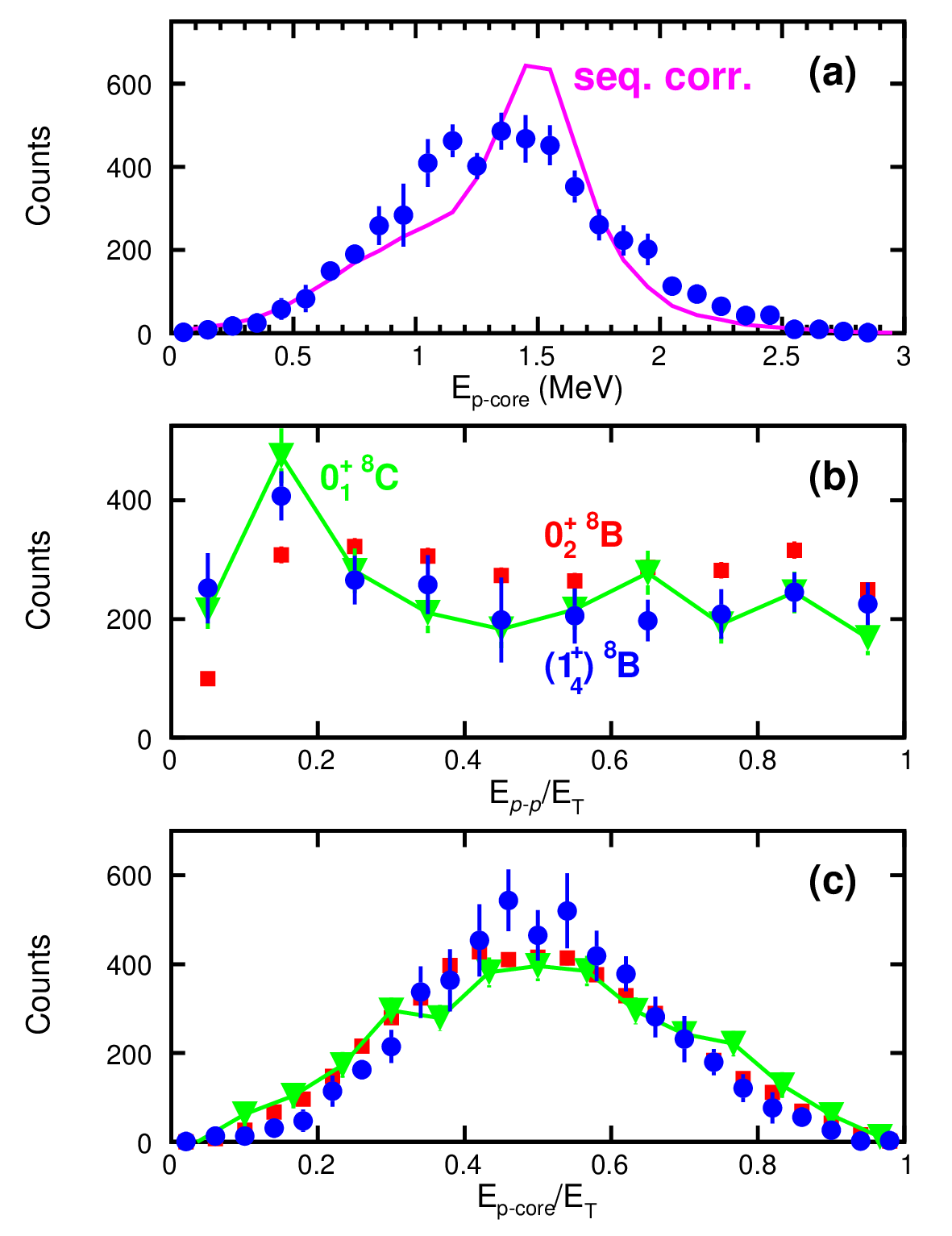}
\caption{(a) Data points show the background-subtracted distribution of the $p$-core relative energy in the 2$p$ decay of the 8.4-MeV state.  The magenta curve shows the expectation for sequential 2$p$ decay through the $^7$Be(5/2$^-_2$) intermediate state. 
(b,c)  Decay correlations for this 8.4-MeV state are compared to those for $^8$C(g.s.) and its isobaric analog in $^8$B. To help differentiating these data, the green data points are connected by the green lines.
} 
\label{fig:corr_2pLi6}
\end{figure}

Figures~\ref{fig:corr_2pLi6}(b) and \ref{fig:corr_2pLi6}(c) show the decay correlations for the three states of interest in terms of the fraction of the decay energy in the $p$-$p$ and $p$-core sub-systems. (The core here refers to the daughter, i.e., $^6$Li[1$^+$, T = 0], $^6$Li[0$^+$, T = 1], and $^6$Be[0$^+$, T = 1], from bottom to top in Fig.~\ref{fig:level_2pLi6}.)
The results for $^8$C(0$^+_1$) and $^8$B(0$^+_2$) (red and green data points) are from \cite{Charity:2011,Brown:2014}. Background-subtracted correlations for the 8.4-MeV state were extracted in Ref.~\cite{CharityLongArxiv} from the two invariant-mass data sets 
(with the $^9$C and $^{13}$O beams) and found to be consistent. The results shown by the blue data points represents the average of the two data sets with the error bars indicating their spread.

The correlations for the three 2$p$ emitters are strikingly similar suggesting that the new state is also a prompt 2$p$ emitter. All three $E_{p\textrm {-core}}/E_{T}$ distributions, Fig.~\ref{fig:corr_2pLi6}(c), are peaked close to 0.5 where the two protons are emitted with the same energy. This is a characteristic feature of prompt 2$p$ decay as it maximizes the product of their barrier penetration factors. This feature is observed in the correlations of all other ground-state 2$p$ emitters \cite{Egorova:2012,Webb:2019a,Brown:2014b}. Having a maximum at $E_{p\textrm {-core}}/E_{T}$=0.5, also implies that the two protons experience the same barrier and therefore are emitted with the same $\ell$ value. This indicates that the 8.4-MeV state has positive parity consistent with the knockout of a $p$-wave proton in the $^9$C data set.

Substantial sequential decay of the newly found state, through the known 5/2$^-_2$ intermediate resonance in $^7$Be, can be more definitively excluded from the invariant-mass distribution of the two $p$-$^6$Li sub events in each 2$p$-$^6$Li event. The data points in Fig.~\ref{fig:corr_2pLi6}(a) show this distribution. The magenta curve shows the expectation for sequential 2$p$ decay obtained using the $R$-matrix formalism \cite{Lane:1958}. This distribution exhibits a peak at $E_{p\textrm{-}^{6}{\rm Li}}\approx$1.5~MeV (from the second sequential step) which is not observed in the data. This peak would also exist in the $E_{p\textrm {-core}}/E_{T}$ distribution shown in Fig.~\ref{fig:corr_2pLi6}(c), but as the $^{8}$B peak is wide, dividing by $E_{T}$ smears it out.

For the data set with the $^9$C beam, an invariant-mass peak for $p$+$^3$He+$\alpha$ exit channel was found with $E^*$=8.23(18)~MeV and $\Gamma$=1.30(33) \cite{CharityLongArxiv} which are close to the values for the 8.4-MeV peak.  
 This could possibly represents a second decay branch of this 2$p$ emitter.  If so, the $p$+$^3$He+$\alpha$ branch has a yield 1.2(4) times that of the 2$p$ branch \cite{CharityLongArxiv}. Correlations indicate that the $p$+$^3$He+$\alpha$ exit channel proceeds by an initial $^3$He+$^5$Li decay and possibly a smaller contribution from $p$+$^7$Be(5/2$^-_1$) decay \cite{CharityLongArxiv}.

\textit{Theory. ---} We have employed the symmetry-adapted no-core shell model (SA-NCSM) \cite{DytrychLDRWRBB20,LauneyDD16}  with the NNLO$_{\rm opt}$ chiral potential \cite{Ekstrom:2013} to calculate the excitation spectrum of $^8$B and the intrinsic structure of the mass $A=6$--$8$ 
states of interest. SA-NCSM is an \emph{ab initio} approach that has been used to successfully predict energy spectra \cite{Sargsyan_A8,heller2022new}, electric quadrupole moments \cite{LauneyMD_ARNPS21} and transition strengths \cite{Henderson:2017dqc, Ruotsalainen19,PhysRevC.100.014322}, beta decays, clustering features \cite{Sargsyan_A8,DreyfussLESBDD20,DreyfussLTDBDB16}, and reaction dynamics \cite{Mercenne:2022,LauneyMD_ARNPS21,BurrowsSumRules}, for nuclei up to the calcium region. SA-NCSM yields identical results to the traditional no-core shell model \cite{NavratilVB00,BarrettNV13} for a given inter-nucleon interaction.

\textit{Discussion. ---} Based on the SA-NCSM calculations, candidate levels that can be produced via $p$-wave proton knockout from the $^9$C beam (0$^+$, 1$^+$, 2$^+$, and 3$^+$) and located in excitation close to the newly observed 2$p$ emitter are the 1$^+_4$, 2$^+_4$ and the 3$^+_3$ states \cite{CharityLongArxiv}. These calculations largely reproduce the known levels at lower excitation. The single-proton decay energies of these states to the ground (3/2$^-$) and first excited (1/2$^-$) states in $^7$Be are more than twice the height of the Coulomb plus centrifugal barrier. In order for a 2$p$ decay branch to complete, the 1$p$ channels must be associated with very small spectroscopic factors. The SA-NCSM  spectroscopic factors for the candidate states to these low-lying states in $^7$Be (Table~\ref{tbl:SP}) are indeed quite small ($<8\%)$.

The model calculations indicate that the three candidates have significant spectroscopic factors for proton decay to either of the two higher-lying adjacent 5/2$^-$ states in $^7$Be \cite{CharityLongArxiv}. Experimentally, the 5/2$^-_1$ state decays almost exclusively to the $^3$He+$\alpha$ channel \cite{Tombrello:1963,Spiger:1967} while the 5/2$^-_2$ state decays predominantly to the $p$+$^6$Li(g.s.) channel \cite{McCray:1963} which is consistent with the SA-NCSM calculations \cite{CharityLongArxiv} and recent Gamow shell-model results \cite{Fernandez:2023}.

The 2$^+_4$ and 3$^+_3$ states have large spectroscopic factors for decay to the 5/2$^{-}_{1}$ level of $^7$Be (1.08 and 0.81, respectively) and thus the decay of these levels should populate the $p$+$^3$He+$\alpha$ final state and possibly could help explain the invariant-mass peak observed for that channel at the same value of E$^*$. The 1$^+_4$ level is predicted to have a significant spectroscopic strength for decay to the 5/2$^{-}_{2}$ proton resonance in $^7$Be and thus could populate the 2$p$+$^6$Li final exit channel via a sequential 2$p$ decay.  However, SA-NCSM calculations show that the proton width through the 5/2$^{-}_{2}$, while not negligible, is significantly smaller than the $2p$ width. This difference arises from the fact that prompt $2p$ decay can emit two $s$-wave isovector paired protons (practically inaccessible to $2^+_4$ and $3^+_3$), whereas a sequential 2$p$ decay involves a $p$-wave proton emission. This corroborates the experimental results, namely, such a sequential decay is not seen in the experimental correlations [Fig.~\ref{fig:corr_2pLi6}(a)]. This suggests that the decay of 1$^+_4$ to $^6$Li is similar to the ground-state 2$p$ emitters $^6$Be \cite{Egorova:2012} and $^8$C \cite{Charity:2011} where the $s$-wave proton pair channel is accessible and in competition with $p$-wave sequential $2p$ decay.

\begin{figure}[!htb]
\includegraphics[width=1\linewidth]{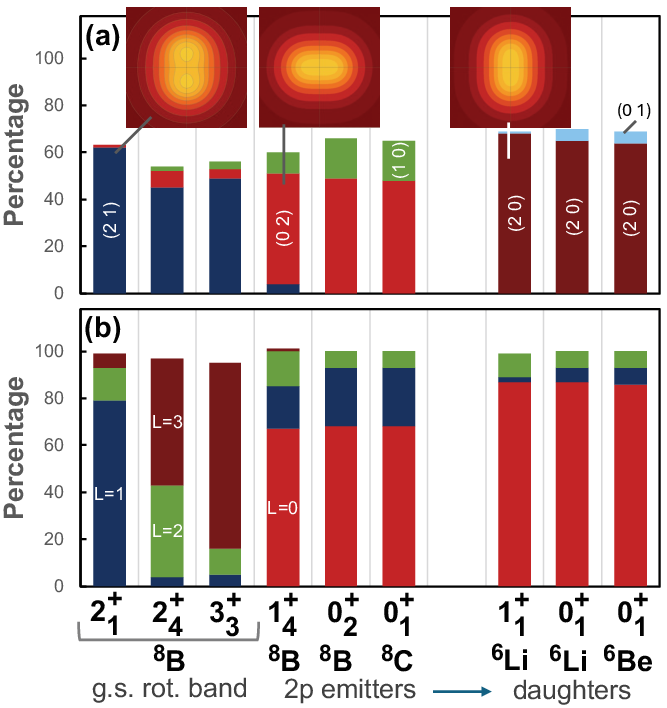}
\caption{
Probability amplitudes of the strongest configurations predicted by the SA-NCSM for states of interest in this work, given by (a) the distribution of deformation, specified by the 
$\lambda$ and $\mu$ 
SU(3) quantum numbers of the SA basis; specifically, $(0 \, 0)$, $\lambda>\mu$, and $\lambda < \mu$, describe spherical, prolate, and oblate deformations, respectively, and (b) the distribution of $L$.
Also shown in (a) are the one-body density profiles in the intrinsic frame ($z$ vs $r_{xy}=\sqrt{x^2+y^2}$, where $z$ is the symmetry axis) for the most dominant configurations.
}
\label{fig:percentage}
\end{figure}

When we consider the dominant deformation configurations available from the SA-NCSM calculations, it becomes evident that the 1$^+_4$ state does in fact have a high degree of similarly to the other two $A$=8 2$p$ emitters [$^8$B(0$^+_2$) and $^8$C(0$^+_1$)] and hence is the best candidate for the new 2$p$ decay (Fig.~\ref{fig:percentage}a). The wavefunctions of the parent 2$p$ emitters are dominated by an oblate configuration while those of their daughter states [$^6$Li(1$^+_1$), $^6$Li(0$^+_1$), and $^6$Be(0$^+_1$)] are dominated by a prolate configuration. Thus the protons are emitted from the equatorial regions.

In addition, to have a 2$p$ decay unaffected by potential intermediates, a significant overlap between the total orbital angular momentum 
($L$) of the initial and final states should exist [Fig.~\ref{fig:percentage}(b)].   
With the $^6$Li ground state being predominantly $L$=0, and if we assume both protons are emitted from $p$ or $s$ orbitals, 
only $^8$B states with $L$=0, 1, or 2 components should be considered as possible candidates.
 The 3$^+_3$ assignment can be ruled out as it is predominantly $L$=3 (79\%). The 2$^+_4$ state 
has a significant $L$=2 projection (39\%) so it a plausible assignment. The best candidate remains the 1$^+_4$ level which is predominantly $L$=0 (87\%), with a moderate $L$=2 projection (11\%).

The 2$p$ decay of the $T$=2 members of the triplet are $0^+\rightarrow 0^+$ transitions and thus the two emitted protons have zero total angular momentum. With the 1$^+_4$ assignment, the new case is a $1^+\rightarrow 1^+$ transition allowing for similar angular-momentum coupling of the two protons thus leading to similar correlations. This is not true for the other candidate spin assignments where the spin couplings of the protons would be completely different.  However, the correlations in the triplet are not expected to be completely  identical as 
their decay energies differ: $Q_{2p}$=1.3~MeV (0$^+_2$ $^8$B), 2.1~MeV (0$^+_1$ $^8$C), and 2.7~MeV (1$^+_4$ $^8$B).  Subtle differences could also result from the different relative locations and widths of the intermediate states.  

The possibility that the 1$^+_4$ state has a significant decay branch to the $p$+$^3$He+$\alpha$ exit channel is not supported by the SA-NCSM calculations for which the $^3$He+$^5$Li and $p$+$^7$Be(5/2$^-_1$) decay  branches are minor \cite{CharityLongArxiv}.

The same structure should be present in the levels of the mirror nuclei.
The mirror 1$^+_4$ level in $^8$Li (Fig.~\ref{fig:isoDia}) has been studied via  $\beta$-delay triton emission of $^8$He.
Borge \textit{et al.} \cite{Borge:1993} measured the $t$ energy spectrum, which was fit with a $t$+$^5$He$\rightarrow t$+$n$+$\alpha$ sequential decay model, giving an excitation energy of 9.3(1)~MeV.
This decay path is the mirror of the $^3$He+$^5$Li path which dominates the state observed in the $p$+$^{3}$He+$\alpha$ channel observed at almost the same energy. This adds support to the contention that we observed the $^3$He+$^5$Li decay branch of $^8$B(1$^+_4$), however, there  are other interpretations of the $t$-energy spectrum.
Barker's analysis, allowing for the possibility of more levels and channels, deduced E$^*$=9.02-9.92~MeV \cite{Barker:1996}.  Grigorenko \textit{et al.} using a three-body $\alpha$+$t$+$n$ model showed that the triton spectrum could be fit without introducing correlations, i.e. with a simple phase-space approach, and found E$^*$=8.5-8.8~MeV \cite{Grigorenko:1996}.  Clearly the deduced excitation energy, of the 1$^+_4$ level in the mirror $^8$Li, is model dependent.  The SA-NCSM and the Greens function Monte Carlo calculations \cite{Wiringa:2000} favor excitation energies at the lower end of the range of these three analyses \cite{CharityLongArxiv}.

All three analyses of the triton spectrum agree that the $B$(GT) value for the 1$^+_4$ state is very large, in fact one of the largest known. This indicates that the giant Gamow Teller resonance is concentrated in this 1$^+_4$ state and that this state has a structure similar to $^8$He(g.s.) (cf. Ref.~\cite{DytrychLDRWRBB20} for the $^8$He deformation). Grigorenko \textit{et al.} refer to it as a ``Halo Analog State'' \cite{Grigorenko:1996}. This title is also appropriate for our triplet of 2$p$ emitters and for their daughter states which are analogs of $^6$He(g.s.). While the overall similarity among the parent (daughter) halo analog states is striking, their detailed structure and asymptotics differ, largely due to the Coulomb force.

\textit{Conclusions. ---}
The nature of the $E^*$=8.4-MeV resonance state in $^8$B observed in \cite{CharityLongArxiv} that decays to the 2$p$+$^6$Li channel
has been established and shown to have an unexpected resemblance to neighboring resonances. Momentum correlations between the decay fragments indicate that this is a prompt 2$p$ emitting state. Based on comparison to \textit{ab initio} symmetry-adapted no-core shell-model calculations, this level is assigned as the forth 1$^+$ state which has a spatial structure very similar to $^8$C and its isobaric analog state in $^8$B ($T$=2 analog states).  Indeed, this newly found state and the two just mentioned constitute a triplet of kindred 2$p$ emitters with similar correlations between the decay products and which, via these decays, remove one unit of isospin, leave the spin and parity unchanged, and converts an oblate parent into a prolate daughter.

 \begin{acknowledgments}
This work is supported by the U.S. Department of Energy, Office of Science, Office of Nuclear Physics under Awards No. DE-FG02-87ER-40316, DE-SC0023532, and under the FRIB Theory Alliance award DE-SC0013617. This work is also supported in part by the National Nuclear Security Administration through the Center for Excellence in Nuclear Training and University Based Research (CENTAUR) under grant number DE-NA-0004150. This work benefited from high performance computational resources provided by LSU (www.hpc.lsu.edu), the National Energy Research Scientific Computing Center (NERSC), a U.S. Department of Energy Office of Science User Facility at Lawrence Berkeley National Laboratory operated under Contract No. DE-AC02-05CH11231, as well as the Frontera computing project at the Texas Advanced Computing Center, made possible by National Science Foundation award OAC-1818253.

  \end{acknowledgments}


\bibliography{extract_B8short}
\newpage
\appendix

\textbf{End Matter}

Spectroscopic factors and decay widths for states of interest, calculated with the SA-NCSM  with continuum~\cite{LauneyMD_ARNPS21}, are listed in Table~\ref{tbl:SP}.
\begin{table}[ht]
\caption{\label{tbl:SP} Spectroscopic factors (SF) calculated by the SA-NCSM with NNLO$_{\rm opt}$ for the overlap of $^8$B 1$^+$, 2$^+$, and 3$^+$ states of interest with the  proton plus various $^7$Be states, along with the corresponding partial decay widths for the largest SFs,
along with $2p$ widths.
For details, see the companion paper~\cite{CharityLongArxiv}.}



\begin{ruledtabular}
\begin{tabular}{ccccc}

 Channel&   1$^+_4$  & 2$^+_4$ & 3$^+_3$  \\
 \colrule
 &\multicolumn{3}{c}{SF}\\
$p$+$^7$Be(1/2$^-$)  & 0.03  & 0.06  & $<$0.01  \\
$p$+$^7$Be(3/2$^-$)  & 0.02  & 0.02 & 0.08 & \\
$p$+$^7$Be(5/2$^-_1$) & 0.01  & 1.08  & 0.81 \\
$p$+$^7$Be(5/2$^-_2$)  & 0.69  & 0.08  & 0.11  \\
$p$+$^7$Be(7/2$^-$)  & 0.001  & $<$0.001  & 0.25 \\
\colrule
 &\multicolumn{3}{c}{$\Gamma$ (MeV)}\\
 $p$+$^7$Be(5/2$^-_1$) &  & 0.8(4) &  0.6(3)\\
 $p$+$^7$Be(5/2$^-_2$) & 0.3(2)\\
 $2p$+$^6$Li(1$^+_1$)  & 0.8(2) & 0.04(2) &  0.02(1)
\end{tabular}
\end{ruledtabular}

\end{table}
\end{document}